\documentclass[pre,twocolumn,showpacs,showkeys,amsfonts,amssymb,amsmath,eqsecnum,letterpaper,nofootinbib,floatfix]{revtex4}

\usepackage{graphicx}  
\begin{document}   
\newlength{\GraphicsWidth}
\setlength{\GraphicsWidth}{8cm}     

\newcommand\comment[1]{\textsc{{#1}}}

\renewcommand\r{\mathbf{r}}

\title{%
A density functional theory study of electric potential saturation:
planar geometry
}
\author{Gabriel T\'ellez}
\email{gtellez@uniandes.edu.co}
\affiliation{Departamento de F\'{\i}sica, Universidad de Los Andes,
A.A.~4976, Bogot\'a, Colombia}
\author{Emmanuel Trizac}
\email{Emmanuel.Trizac@th.u-psud.fr}
\affiliation{Laboratoire de Physique Th\'eorique (Unit\'e Mixte de
Recherche UMR 8627 du CNRS), B\^atiment 210, Universit\'e de
Paris-Sud, 91405 Orsay Cedex, France}

\begin{abstract}                              
We investigate the possibility of electrostatic potential saturation,
which may lead to the phenomenon of effective charge saturation.
The system under study is a uniformly charged infinite plane 
immersed in an arbitrary electrolyte made up of several micro-species.
To describe the electric double layer,
we use a generic density functional theory  
in which the local micro-ionic density profiles
are arbitrary functions of the local electrostatic potential. A necessary 
and sufficient condition is obtained for saturation, whereby
the electrostatic potential created by the plane becomes independent 
of its bare charge, provided the latter is large enough.
\end{abstract}


\maketitle


\section{Introduction}

The importance of effective charge in colloidal or polyelectrolyte
suspensions has been recognized for some
time~\cite{Manning,Alexander,Belloni,Hansen,Levin,Quesada,Trizac1,NetzOrland-charge-renorm}.
The electric potential far from a colloid immersed in an electrolyte
defining the inverse screening length $\kappa$ takes the same form as
the solution of the linearized Debye--H\"uckel equation ---say $l_B
Z_{\text{eff}} e^{-\kappa(r-R)} / ((1+\kappa R)r)$ for a charged
sphere of radius $R$--- but with a prefactor $Z_{\text{eff}}$ which is
different from the bare charge $Z_{\text{bare}}$ of the colloid. In
the previous expression, $r$ is the distance from the center of the
sphere of interest, and $l_B = q^2/(\varepsilon k_BT)$ is the Bjerrum
length, defined from the elementary charge $q$ and the permittivity
$\varepsilon$ of the solvent (the molecular structure of which is
neglected).  An interesting feature that occurs in the framework of
(non-linear) Poisson--Boltzmann theory \cite{Israelachvili} is that
for highly charged colloids, the effective charge $Z_{\text{eff}}$
saturates to a finite value $Z_{\text{sat}}$
\cite{Trizac1,JPA}. Interestingly, this saturation value is an upper
bound for effective charges found within more refined approaches that
incorporate the micro-ionic correlations neglected in
Poisson-Boltzmann theory \cite{Groot}. We will come back to the
validity of the latter approach in the concluding section.

Since Poisson--Boltzmann is a mean field theory, there has been
several proposals to go beyond that approximation and try to include
effects like micro-ions excluded volume~\cite{Andelman} or
electrostatic correlations between micro-ions~\cite{Barbosa}. A
natural question arises. Under which conditions a given theory will
account for the saturation phenomenon? We will not give here a
definite answer to this question in its full generality. However we
will consider the special case of local theories ---theories which
provide a local relationship between the density of micro-ions and the
electric potential--- and give a necessary and sufficient condition
under which the electrostatic potential of an infinite charged plane
immersed in an electrolyte will saturate when the bare surface charge
density of the plane diverges. As explained below, the phenomenon of
potential saturation is slightly more general than that of effective
charge saturation: effective charge saturation implies potential
saturation, whereas potential saturation does not necessarily allows
to define an effective charge.

This paper is organized as follows. We will present in
Sec.~\ref{sec:framework} the framework and the class of theories that
will be considered. In Sec.~\ref{sec:integration}, we will formally
integrate the generalization of Poisson--Boltzmann equation that these
local theories yield.  We will discuss in
Sec.~\ref{sec:charge-saturation} the necessary conditions to have the
potential saturation effect. Finally, in Sec.~\ref{sec:examples}, we
will apply our results to some recent theories~\cite{Andelman,Barbosa}
that have been proposed to go beyond the description of
Poisson--Boltzmann theory.


\section{Framework}
\label{sec:framework}

Let us consider an infinite plane located at $x=0$ with a surface
charge density $\sigma$. The half-space $x>0$ is filled with an
electrolyte made up of several species of micro-ions with charges
$q_{\alpha}$ and local densities $n_{\alpha}(x)$. We use Greek
indexes for the different species of micro-ions. The system is in thermal
equilibrium at an inverse temperature $\beta=(k_B T)^{-1}$. Without
loss of generality we suppose $\sigma>0$.

The average electric potential $\psi(x)$ at a distance $x$ from the
charged plane is related to the densities of micro-ions by Poisson
equation
\begin{equation}
\label{eq:Poisson}
\psi''(x)=- \frac{4\pi}{\varepsilon} \rho(x).
\end{equation}
The prime denotes differentiation with
respect to $x$ and $\rho(x)=\sum_{\alpha} q_{\alpha}
n_{\alpha}(x)+\sigma \delta(x)$ is the total charge
density. Furthermore the electric potential satisfies the boundary
conditions
\begin{subequations}
\label{eq:BC-psi}
\begin{eqnarray}
\label{eq:BC-psi-x=0}
\psi'(0^+)&=&-4\pi\sigma/\varepsilon\\
\label{eq:BC-psi-x=infinity}
\psi(\infty)=0, &&\psi'(\infty)=0,\ ~ \ \psi''(\infty)=0
\end{eqnarray}
\end{subequations}
Suppose now that an approximate local theory is provided. This theory
gives local relations between the densities and the electric potential
\begin{equation}
\label{eq:local-relation-n-psi}
n_{\alpha}(x)=g_{\alpha}[q_{\alpha} \psi(x)]
\end{equation}
which leads to a closed equation for the electric potential when we
replace Eq.~(\ref{eq:local-relation-n-psi}) into
Eq.~(\ref{eq:Poisson})
\begin{equation}
\label{eq:Poisson-Boltzmann-general}
\psi''(x)=-\frac{4\pi}{\varepsilon}\sum_{\alpha} q_{\alpha} g_{\alpha}[q_{\alpha}\psi(x)], 
\quad \hbox{for } x>0
\end{equation}
When $g_{\alpha}(u)=n_{\alpha}^b\exp(-\beta u)$, with $n_{\alpha}^b$
the bulk density of the species $\alpha$, we recover Poisson--Boltzmann
equation.  Equation~(\ref{eq:Poisson-Boltzmann-general}) may then be
considered as a generalized
Poisson--Boltzmann equation.

The local relation~(\ref{eq:local-relation-n-psi}) between the density
and the electric potential may be obtained in the framework of the
density functional theory (DFT) by a local density approximation
(LDA). In this framework, the free energy functional is given by
\begin{eqnarray}
\label{eq:DFT-LDA}
\mathcal{F}(\{n_{\alpha}\})&=&
\int_{0}^{\infty} f(\{n_{\alpha}(x)\})\,dx
\\
&+&\frac{1}{2} \int_0^{\infty}\int_0^{\infty}
\rho(x) \mathcal{G}(x,x') \rho(x')\,dx\,dx'
\nonumber
\end{eqnarray}
where $\mathcal{G}$ is ($-4\pi/\varepsilon$) times the one-dimensional
Laplacian Green function with the appropriate (Neumann) boundary
conditions. We have $\psi(x)=\int_0^{\infty} G(x,x')\rho(x')\,dx'$.

The minimization of the functional~(\ref{eq:DFT-LDA}) subject to the
conservation of the total number of particles (controlled by Lagrange
multipliers $\mu_{\alpha}$, the chemical potentials of micro-species)
gives
\begin{equation}
\label{eq:minimization}
q_{\alpha} \psi = 
-\frac{\partial f(\{n_{\gamma}\})}{\partial n_{\alpha}}
+\mu_{\alpha}
\end{equation}
We assume that upon inverting 
these relations, one obtains a set of relations 
of the
form~(\ref{eq:local-relation-n-psi}) between each
density and the electric potential. Note that
Eq.~(\ref{eq:local-relation-n-psi}) is really a set of equations
giving, for each $\alpha$, an explicit relation between
$n_{\alpha}(x)$ and the electric potential $\psi(x)$. In particular
each relation involves only one density $n_{\alpha}(x)$. It can,
however, include the charges $q_{\gamma}$ and chemical potentials
$\mu_{\gamma}$ of the other particles~\cite{Andelman} but not the
other local densities $n_{\gamma}(x)$.

Poisson--Boltzmann theory is recovered when the local part of the free
energy is given by the ideal gas contribution
$f(\{n_{\alpha}\})=f_{\text{id}}(\{n_{\alpha}\})=\beta^{-1}
\sum_{\alpha} n_{\alpha} [\ln(n_{\alpha} \Lambda)-1]$, where $\Lambda$
is an irrelevant length (the de Broglie wavelength).

The relations~(\ref{eq:local-relation-n-psi}) should obey a certain
number of physical constraints, for instance the local charge density
$\rho(x)$ should vanish when $x\to\infty$. We will impose the
following constraint to the functions $g_{\alpha}$ appearing in
Eq.~(\ref{eq:local-relation-n-psi}) 
\begin{equation}
\label{eq:hypothese-stabilite}
\sum_{\alpha} q_{\alpha} g_{\alpha}[q_{\alpha}\psi(x)] \psi'(x)
>0
\end{equation}
This condition will be used several times in the subsequent
analysis. In the appendix~\ref{sec:appendix} we show that, in the
framework of the DFT, this condition is a consequence of the stability
of the system, namely that $\partial^2 f(\{n_{\alpha}\})/\partial
n_{\alpha} \partial n_{\gamma}$ is positive definite.


\section{Formal integration of the generalized Poisson--Boltzmann
equation}
\label{sec:integration}

Multiplying Eq.~(\ref{eq:Poisson-Boltzmann-general}) by $\psi'(x)$, we
get
\begin{equation}
\frac{d}{dx}\left[\left(\psi'(x)\right)^2\right]= -\frac{8\pi}{\varepsilon} 
\sum_{\alpha}
q_{\alpha} g_{\alpha}(q_{\alpha}\psi) \frac{d\psi}{dx}
\end{equation}
which allows for a first integration
\begin{equation}
\label{eq:psi-prime}
\psi'(x)=-\sqrt{-\frac{8\pi}{\varepsilon}
\sum_{\alpha} G_{\alpha}(q_{\alpha} \psi(x))}
\end{equation}
where we defined 
\begin{equation}
\label{eq:def-G_alpha}
G_{\alpha}(v)=\int_{0}^{v} g_{\alpha}(u)\,du
\end{equation}
and we have used the boundary condition~(\ref{eq:BC-psi-x=infinity})
at $x\to\infty$. The choice of the minus sign in the r.h.s.~of
Eq.~(\ref{eq:psi-prime}) is dictated by the fact that $\sigma>0$.

Note that condition~(\ref{eq:hypothese-stabilite}) ensures that the
term under the square root sign in Eq.~(\ref{eq:psi-prime}) is
positive. Indeed, (\ref{eq:hypothese-stabilite}) implies
that $d(\psi'(x)^2)/dx<0$, therefore $\psi'(x)^2$ is a decreasing
function and never vanishes for $x$ finite. Then, $\psi'(x)$ never
changes sign and since $\sigma>0$, $\psi'(x)<0$ and we conclude that
$\psi'(x)$ is monotonous and increasing. The electric field
$-\psi'(x)$ is monotonously decreasing, always positive and
non-vanishing for $x$ finite. This remark is in fact a consequence for
this particular geometry of the general proof on the absence of
over-charging whereby electric double layers are described by a local
density functional theory of the form (\ref{eq:DFT-LDA})
~\cite{Trizac-charge-like}.

Upon formally integrating 
Eq.~(\ref{eq:psi-prime}), we obtain
\begin{equation}
\label{eq:x-psi}
F(\psi)= - (x+x_0)
\end{equation}
with $x_0$ a constant of integration and the indefinite integral
\begin{equation}
\label{eq:function-F}
F(\psi)=\int^{\psi} 
\frac{\sqrt{\varepsilon} \, d\phi}{
\sqrt{-8\pi
\sum_{\alpha} G_{\alpha}(q_{\alpha} \phi)}
}
\end{equation}
The solution for the electric potential is given by inverting
relation~(\ref{eq:x-psi})
\begin{equation}
\psi(x)=F^{-1}[-(x+x_0)]
\end{equation}

The function $F(\psi)$ introduced in Eq.~(\ref{eq:function-F}) has a
few useful properties. It is the integral of a positive quantity, so that
it is a strictly increasing function of $\psi$ for
$\psi\in\left[0,\infty\right[$%
%
%
\footnote{
\label{footnote1}
We suppose that the function $F(\psi)$ exists for all values of
$\psi>0$. This will be the case if all the functions
$G_{\alpha}(q_{\alpha}\psi)$ are defined for any value of $\psi$.}
%
%
that can be inverted: $F^{-1}$ exists
and it is also an increasing function. As a consequence $\psi(x)=F^{-1}[-(x+x_0)]$
is a decreasing function of $x$.

The constant of integration $x_0$ is related to the surface charge
density $\sigma$ of the plane and is determined by the boundary
condition~(\ref{eq:BC-psi-x=0}). An interesting feature is that this
constant of integration comes as an additive offset for the position
$x$. A change in $\sigma$ (therefore in the constant of integration
$x_0$) results in a translation of the curve $\psi(x)$--$x$ along
the $x$-axis.  This allows a graphical determination of $x_0$:
plotting the function $F^{-1}(-x)$, the origin $x=0$ of the $x$-axis 
should be such that
$dF^{-1}(-x)/dx=-4\pi\sigma/\varepsilon$ at this new origin, thus satisfying the
boundary condition~(\ref{eq:BC-psi-x=0}). Again, the stability
condition~(\ref{eq:hypothese-stabilite}) ensures that there is a
unique solution for $x_0$ since $\psi'(x)$ is monotonous.


\section{Electrostatic potential saturation}
\label{sec:charge-saturation}

We will say that there is a saturation of the electrostatic potential
if it is possible to have $\sigma\to+\infty$ with a finite solution
for $\psi(x)$ for all $x\neq0$\footnote{It is important to exclude
$x=0$. Indeed, the limit $\sigma\to+\infty$ implies that
$\psi'(0)=-\infty$ but since both $\psi $ and $\psi'$ are monotonous
and $F(\psi)$ is supposed to exist for all values of $\psi>0$, we also
have $\psi(0)=+\infty$. The case of $\psi(0)>0$ finite while
$\psi'(0)=+\infty$ is ruled out by the conditions stated in
footnote~\ref{footnote1}: if this was the case it would mean that the
integral~(\ref{eq:function-F}) defining $F(\psi)$ would not be defined
for values of $\psi>\psi(0)$ which implies that
$\sum_{\alpha}G_{\alpha}(q_{\alpha} \psi_{\alpha})$ diverges to
$-\infty$ for $\psi=\psi(0)$ and is not defined for $\psi>\psi(0)$.
}. We emphasize that the notion of potential saturation is more
general than that of effective charge saturation: to define an
effective charge, one needs to show that the ionic profiles behave far
from the wall as they would within a {\em linear} theory such as
Debye--H\"uckel.  The effective charge is then defined from the far
field created by the charged object, as that required within a linear
theory to obtain the same potential at large distances.  Our analysis
does not require such a limitation.  However, in the case where the
local theory is formulated in the framework of the DFT we show in
appendix~\ref{sec:appendix-B} that, far from the charged wall, the
theory reproduces the far field of the linear Debye--H\"uckel theory,
provided that the Hessian matrix $\left(\partial^2 f/\partial
n_{\alpha}\partial n_{\gamma}\right)$ is positive
definite.

The function $F(\psi)$ is strictly increasing in the interval
$[0,+\infty[$. Therefore there are only two possibilities for the
behavior of $F(\psi)$ when $\psi\to+\infty$ which, as we show
below, distinguish between the cases of saturation and
non-saturation.
\begin{subequations}
\label{eq:conditions-saturation-both1}
\begin{equation}
\label{eq:condition-saturation1}
\text{If\ }\lim_{\psi\to+\infty} F(\psi)<+\infty ,\
\text{there is saturation.}
\end{equation}
\begin{equation}
\label{eq:condition-non-saturation1}
\text{If\ }\lim_{\psi\to+\infty} F(\psi)=+\infty ,\ 
\text{there is no saturation.}
\end{equation}
\end{subequations}

\begin{figure}
\includegraphics[width=\GraphicsWidth]{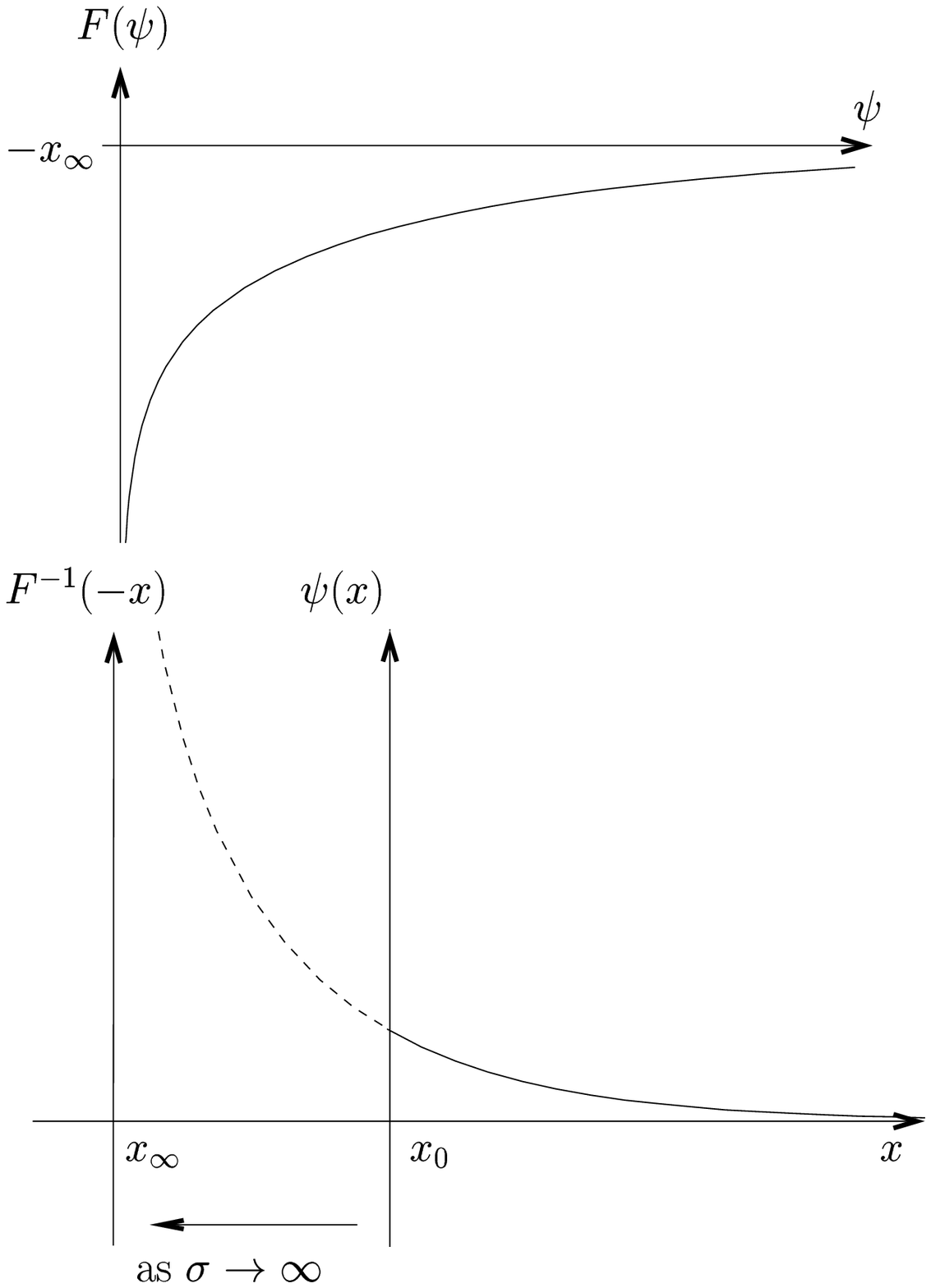}
\caption{
\label{fig:example-saturation}
An example of a theory in which the charge saturation phenomenon is
possible. The upper figure represents the function $F(\psi)$ defined
in Eq.~(\ref{eq:function-F}) and in this example it has a finite limit
when $\psi\to\infty$: $\lim_{\psi\to+\infty}
F(\psi)=-x_{\infty}<+\infty$. The lower figure represents the inverse
function $F^{-1}(-x)$. The plot of the electric potential $\psi(x)$ as
a function of $x$ is simply obtained choosing a new origin for the
$x$-axis, such that $dF^{-1}(-x)/dx=-4\pi\sigma/\varepsilon$ at this new
origin. For a given value of $\sigma$, only the part of the curve at
the right of this origin (full-line) has a physical meaning.
When $\sigma$ increases, this new origin gets translated to the
left to finally reach the position of $x_{\infty}$ when
$\sigma=+\infty$.  }
\end{figure}

Let us consider first the case~(\ref{eq:condition-saturation1}). Let
$\lim_{\psi\to\infty}F(\psi)=-x_{\infty}<+\infty$. The functions
$F(\psi)$ and $\psi(x)$ are sketched in
Figure~\ref{fig:example-saturation}. For any finite value of
$\sigma>0$ the determination of the constant of integration $x_0$ 
gives $x_0>x_{\infty}$ and as $\sigma$ increases, $x_0$ approaches
$x_{\infty}$. The case $\sigma=+\infty$ corresponds to the choice
$x_0=x_{\infty}$. In the graphical way of determining the constant of
integration explained before, this means that the origin of the
$x$-axis is chosen above $x_{\infty}$ if $0<\sigma<+\infty$ and it
approaches $x_{\infty}$ as $\sigma\to+\infty$.  We have clearly the
potential saturation phenomenon since for $\sigma=+\infty$, where the
origin of the $x$-axis is precisely at $x_{\infty}$, we have a finite
solution for the potential $\psi(x)$ for any value of $x>0$.

%
%
%
%

In the second case~(\ref{eq:condition-non-saturation1}), $x_{\infty}$
recedes to $-\infty$.
Again, for $\sigma>0$ finite, the boundary
condition~(\ref{eq:BC-psi-x=0}) yields a value for
$x_0>x_{\infty}=-\infty$, and as $\sigma$ increases, $x_{0}$
decreases. We actually have $\lim_{\sigma\to+\infty} x_0=-\infty$, the
constant of integration should decrease to minus infinity to satisfy
the boundary condition. But since $\psi(x)=F^{-1}(-(x+x_0))$ and
$F^{-1}(+\infty)=+\infty$ then the electric potential $\psi(x)$ is
infinite for all values of $x$. There is no saturation of the potential.

In practice, for a general theory formulated in the framework of a
local DFT, it is not always easy to invert Eq.~(\ref{eq:minimization})
in order to have an explicit
expression~(\ref{eq:local-relation-n-psi}) of the density as a
function of the potential, and even more difficult to compute the
function $F(\psi)$ to study its limit as $\psi\to\infty$.  However,
there is no need to have explicitly the functions $g_{\alpha}$ of
Eq.~(\ref{eq:local-relation-n-psi}) to discuss the possibility of
potential or charge saturation. Indeed, as shown below, we only need
to know the behavior of the functions $g_{\alpha}(q_{\alpha}\psi)$ as
$\psi\to+\infty$.

First we should treat the particular situation where it is possible to
have from Eq.~(\ref{eq:local-relation-n-psi}) or
Eq.~(\ref{eq:minimization}) an infinite potential $\psi\to+\infty$
while all the densities $n_{\alpha}$ are finite. If this is the case,
then the functions $G_{\alpha}(q_{\alpha}\psi)$ behave as $\psi$
---more precisely they are of order $\psi$ which we denote ${\cal
O}(\psi)$--- when $\psi\to+\infty$. Therefore the integrand
$[-\sum_\alpha G_{\alpha}(q_{\alpha} \psi)]^{-1/2}$ in
Eq.~(\ref{eq:function-F}) defining the function $F(\psi)$ behaves as
$\psi^{-1/2}$ ---more precisely $\psi^{-1/2}={\cal O}\{[-\sum_\alpha
G_{\alpha}(q_{\alpha} \psi)]^{-1/2}\}$--- as $\psi\to+\infty$ and it
is not integrable. We have $\lim_{\psi\to+\infty}F(\psi)\to\infty$. We
then conclude that there is no saturation. To summarize, if a local
theory predicts the possibility of having all densities of the
micro-ions finite when $\psi\to\infty$, this theory will consequently
not account for the phenomenon of charge saturation.

Now, in the opposite case where the theory predicts that at least one
density $n_{\alpha}$ ---the density of counterions--- diverges when
$\psi\to\infty$, the existence of the saturation effect depends on how
this density diverges as $\psi\to\infty$. If several densities
diverge, actually what is important is the one that diverges the
fastest as $\psi\to\infty$, say $n_{\gamma}$. Remembering that
$n_{\gamma}(x)=g_{\gamma}(q_{\gamma}\psi)$, suppose that
$g_{\gamma}(q_{\gamma}\psi)$ diverges as a power law $\psi^{\nu}$. We
prove below that if $\nu>1$, then the effect of charge saturation will
be accounted for by the theory, otherwise it will not be. The argument
is simple: suppose $g_{\gamma}(q_{\gamma}\psi)$ behaves as
$\psi^{\nu}$ when $\psi\to\infty$, then $G_{\gamma}(q_{\gamma}\psi)$
behaves as $\psi^{\nu+1}$. The integrand in Eq.~(\ref{eq:function-F})
behaves as $\psi^{-(\nu+1)/2}$ as $\psi\to\infty$ and is therefore
integrable if and only if $(\nu+1)/2>1$, that is $\nu>1$. It is also
clear from this argument that in the cases where the behavior of
$g_{\gamma}(q_{\gamma}\psi)$ is not a power law of $\psi$ but diverges
faster than $\psi^{\nu}$ for some $\nu>1$ there will be a saturation
of the electric potential and in the cases where it diverges as $\psi$
or slower, the electric potential will not saturate.

The remaining cases where $g_{\gamma}(q_{\gamma}\psi)$ diverges faster
than $\psi$ but slower than $\psi^{\nu}$ for any $\nu>1$ can always be
resolved returning to the study of the behavior of
$[-G_{\gamma}(q_{\gamma}\psi)]^{-1/2}$ when $\psi\to\infty$ ---the
dominant part of the integrand of $F(\psi)$ in
Eq.~(\ref{eq:function-F})--- and determine if it is integrable or
not.\footnote{For example, there could be cases like
$[-G_{\gamma}(q_{\gamma}\psi)]^{-1/2}\propto (\psi \ln\psi)^{-1}$ (not
integrable) in which there will not be a saturation of the potential
or cases like $[-G_{\gamma}(q_{\gamma}\psi)]^{-1/2}\propto (\psi
(\ln\psi)^2)^{-1}$ which is integrable and there will be a saturation.
\label{footnote:cas-limites}
}

In the case where the electric potential saturates, the effective
charge at saturation can be expressed in terms of the function
$F(\psi)$ defined by Eq.~(\ref{eq:function-F}). Let us consider the
saturation regime and suppose that one can define an effective charge
at saturation, i.e.~the far field created by the plane has the same
form as the one predicted by the linear Debye--H\"uckel theory (see
appendix~\ref{sec:appendix-B} for details). This means that if
$\psi\to0$ we have $-(4\pi/\varepsilon) \sum_{\alpha} q_{\alpha}
g_{\alpha}(q_{\alpha}\psi) \sim \kappa^{2} \psi$ so that for
$x\to\infty$,
\begin{equation}
\label{eq:psi-sat-DH-forme}
\psi(x)\sim \psi^{\text{sat}} e^{-\kappa x}
\end{equation}
with $\psi^{\text{sat}}$ related to the effective charge at saturation
by the relation
$\psi^{\text{sat}}=4\pi
\sigma_{\text{eff}}^{\text{sat}}/(\kappa\varepsilon)$.
At saturation, in the graphical way of determining the constant of
integration $x_0$, the origin is at $x_{\infty}$. That is
$x_{\infty}=0$ and
\begin{equation}
-x=F(\psi)
\end{equation}
with the particular choice
\begin{equation}
\label{eq:function-F-a-saturation}
F(\psi)=\int^{\psi}_{+\infty} 
\frac{d\phi}{
\sqrt{-\frac{8\pi}{\varepsilon}
\sum_{\alpha} G_{\alpha}(q_{\alpha} \phi)}}
\end{equation}
in the lower limit of integration. To be consistent with
Eq.~(\ref{eq:psi-sat-DH-forme}), we must have $F(\psi)\sim
\kappa^{-1}\ln(\psi/\psi^{\text{sat}})$ for $\psi\to 0$. Therefore we
can extract effective charge at saturation from
\begin{equation}
\psi^{\text{sat}}=\frac{4\pi
\sigma_{\text{eff}}^{\text{sat}}}{\kappa\varepsilon} =\lim_{\psi\to 0}
\left\{\psi \exp\left[-{\kappa F(\psi)}\right]\right\}
\end{equation}


\section{Some examples}
\label{sec:examples}

In this section we apply our results to some local theories
proposed in the literature.
We first illustrate the results of the preceding section with some
simple examples.  We consider two cases (as benchmarks, since
the analytic solution of the problem is known): Poisson--Boltzmann
theory and its linearized counterpart, Debye--H\"uckel theory. 

For Poisson--Boltzmann theory, $g_{\alpha}(q_{\alpha}
\psi)=n_{\alpha}^{b} \exp(-\beta q_{\alpha} \psi)$, then the integrand
in Eq.~(\ref{eq:function-F}) behaves as $\exp(-\beta |q_0| \psi/2)$
when $\psi\to\infty$ and it is integrable ($q_0$ is the charge of the
counterions with highest valency, which by the way have a density that
diverges exponentially faster than the potential $\psi$).  The
function $F(\psi)$ has a finite limit when $\psi\to\infty$; we
therefore recover there the well known fact that effective charge
saturates when $\sigma \to \infty$.

For Debye--H\"uckel theory, $g_{\alpha}(q_{\alpha}
\psi)=n_{\alpha}^{b} (1-\beta q_{\alpha} \psi)$ behaves as $\psi$ when
$\psi\to\infty$ and the integrand in Eq.~(\ref{eq:function-F}) behaves
as $\psi^{-1}$, it is not integrable: there is no charge
saturation. 

Now let us turn our attention to some more interesting
examples. Barbosa \textit{et al.}~\cite{Barbosa} have proposed a local
theory to account for counterions correlations in a one-component
plasma (OCP) description of the electrolyte ---that is a system of
charged counterions immersed in an uniform oppositely charged
background. This theory, referred to as the
Debye--H\"uckel-Hole-Cavity approach, is stable ---it satisfies
condition~(\ref{eq:hypothese-stabilite})--- and the density $n$ of the
counterions is described in the framework~(\ref{eq:DFT-LDA}) of the
DFT. The local free energy density is of the form
$f(n)=f_{\text{id}}(n)+f_{\text{DHHC}}(n)$ where $f_{\text{id}}$ is
the ideal gas part of the free energy and
\begin{widetext}
\begin{equation}
\label{eq:f-DHHC}
\frac{\beta f_{\text{DHHC}}(n)}{n}= \frac{(\kappa a)^2}{4}
- \int_{1}^{\omega} d\bar{\omega}
\left[
\frac{\bar{\omega}^2\Omega(\bar{\omega})^{2/3}}{2(\bar{\omega}^3-1)}
+\frac{\bar{\omega}^3}{(1+\Omega(\bar{\omega})^{1/3})
(\bar{\omega}^2+\bar{\omega}+1)}
\right]
\end{equation}
\end{widetext}
In this equation,
\begin{equation}
\Omega(\bar{\omega})=(\bar{\omega}-1)^3+
\frac{(\kappa a)^{3}}{3 l_B \kappa}
(\bar{\omega}^3-1), ~~
\omega=(1+3 l_B \kappa)^{1/3}
\end{equation}
and $\kappa=\sqrt{4\pi l_B n}$ is the inverse Debye length.
The theory has a parameter $a$ which may be interpreted
as a sort of radius of the micro-ions (non-strictly speaking, because
the counterion density can actually be higher that $a^{-3}$).

The expression~(\ref{eq:f-DHHC}) is quite complicated and there is no
hope to be able to obtain an analytical solution for the function
$F(\psi)$ or even invert the relationship between the potential and
the density obtained from the stationary
equation~(\ref{eq:minimization}). However we only need to investigate
the limit $\psi\to\infty$.

First it is straightforward to see that if the density $n$ is finite,
the potential is finite. Now, if $n\to\infty$ we have
$\beta f_{\text{DHHC}}(n)\sim n^{5/3}\, a^2 (\pi/6)^{2/3}$ and it is
the leading term in $f(n)$. The minimization
equation~(\ref{eq:minimization}) then yields $\beta |q| \psi\sim
n^{2/3} \, a^{2} (5/3) (\pi/6)^{2/3}$ for $n\to\infty$, so that for
$\psi\to\infty$, we have 
\begin{equation}
n\sim \frac{6}{\pi}\left(\frac{3}{5}\right)^{3/2} a^{-3} (\beta|q|
\psi)^{3/2}
\end{equation}
The important fact is that the density of counterions behave as
$\psi^{\nu}$ with $\nu=3/2>1$. This theory therefore leads to a
saturation of the electrostatic potential, and more precisely to a
saturation of effective charge\footnote{To allow for the definition of
an effective charge, Poisson--Boltzmann generalized equation
(\ref{eq:Poisson-Boltzmann-general}) should behave as Helmholtz
equation ($\psi'' \propto \psi$) when $\psi \to 0$.  This is discussed
in appendix~\ref{sec:appendix-B}. }.

On the other hand there are theories that do not account for the
saturation effect. As an example, let us consider the approach
proposed by Borukhov \textit{et al.}~\cite{Andelman}. This is a local
theory which incorporates approximately steric effects due to volume
exclusion between the micro-ions of the electrolyte. The local part of
the free energy is $f(\{n_{\alpha}\})=f_{\text{id}}(\{n_{\alpha}\})+
f_{\text{exc}}(\{n_{\alpha}\})$ with an excess part involving the size
$a$ of micro-ions
\begin{equation}
\label{eq:f-exec}
\beta f_{\text{exc}}(\{n_{\alpha}\})=
\frac{1}{a^3} \left[1-\sum_{\alpha} a^3 n_{\alpha} \right]
\ln\left(1-\sum_{\alpha} a^3 n_{\alpha} \right)
\end{equation}
This theory is also stable, namely $(\partial^2 f/\partial
n_{\alpha}\partial n_{\gamma})$ is positive
definite~\cite{Trizac-Raimbault-charge-like-steric} and it therefore
satisfies the condition~(\ref{eq:hypothese-stabilite}).  From
Eq.~(\ref{eq:f-exec}), there is a higher bound for the value of the
micro-ions densities which is $a^{-3}$: when the counterions density
approaches $a^{-3}$ the electric potential $\psi$ diverges. Therefore,
according to the discussion of last section, this theory does not
allow the electrostatic potential to saturate.

Our approach also predicts, in the case of non-saturation,
if the effective charge will grow faster or slower than the bare
charge. Let us illustrate this point considering again the framework
presented in Ref.~\cite{Andelman}. To simplify the argument, we consider
a two-component electrolyte with counterions/coions of charge $-q$/$q$
and bulk density $n_b$. Poisson's equation then takes the form
\begin{equation}
\beta q \psi''(x) = \kappa^2 \,\frac{\sinh(\beta q \psi)}{1-\zeta +\zeta
\cosh(\beta q\psi)}
\label{eq:andel}
\end{equation}
where $\zeta = 2 n_b a^3$ and $\kappa=\sqrt{8\pi n_b l_B}$.

\begin{figure}
\includegraphics[width=\GraphicsWidth]{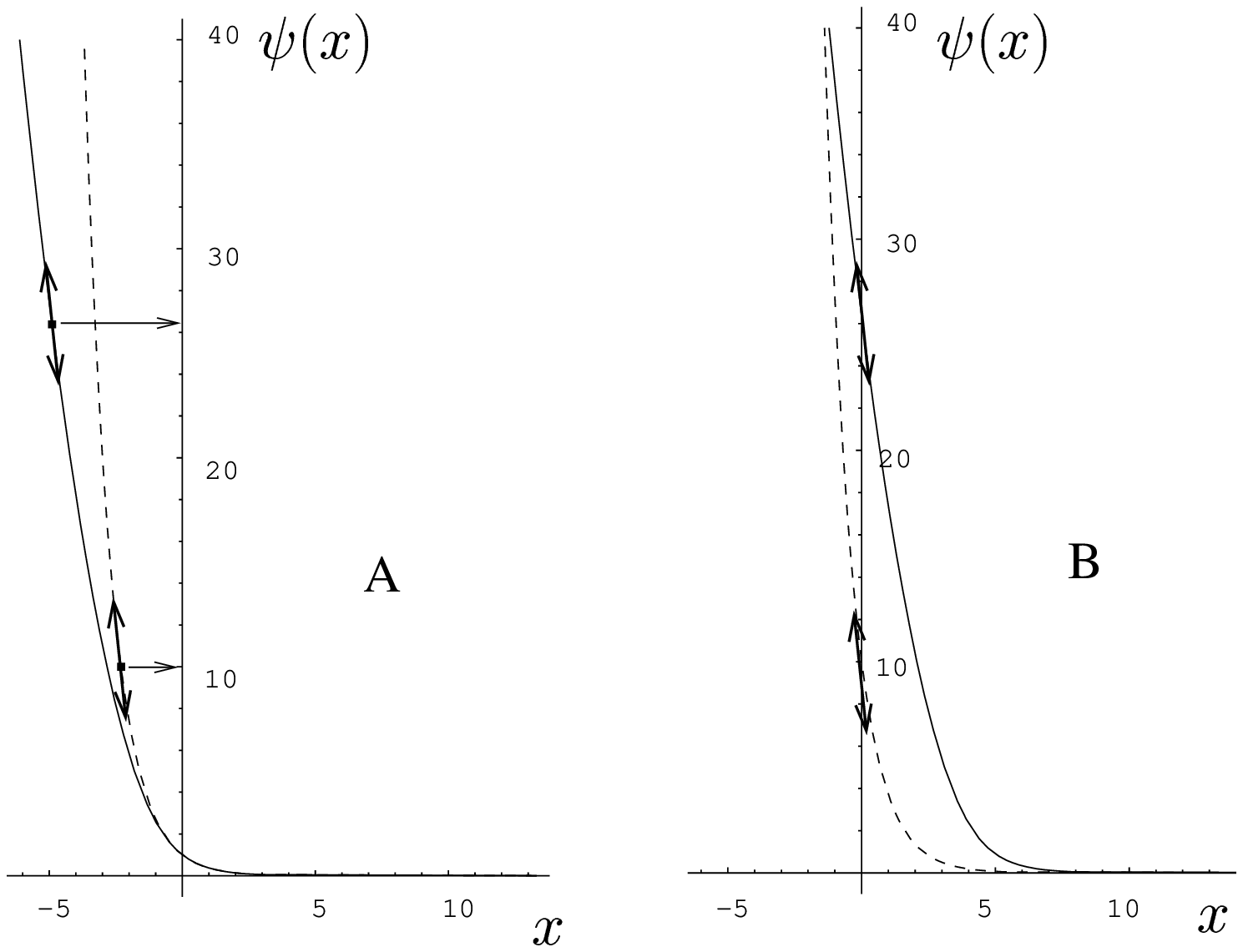}
\caption{
\label{fig:charge-Andelman}
Illustration of a theory (Ref.~\cite{Andelman}) where the effective
charge diverges faster than the bare charge.  The dashed curve
represents the potential $\psi_{\text{DH}}(x)$ predicted by
Debye--H\"uckel theory and the full line represents the potential
$\psi(x)$ put forward in Ref.~\cite{Andelman}. The potential is
measured in units of $kT/q$ and the distance $x$ in units of the Debye
length $\kappa^{-1}$. We have chosen here $a^3 n_b=1/4$, where $n_b$
is the bulk density of ions.  Figure~\ref{fig:charge-Andelman}A
represents a case where the bare charge $\sigma$ is small (actually in
the figure $4\pi\sigma\beta q/(\varepsilon\kappa)=1$). Both curves
have the same behavior as $x\to\infty$, namely,
$\psi(x)=\exp(-x)$. Figure~\ref{fig:charge-Andelman}B represents a
case where $\sigma$ is large ($4\pi\sigma\beta
q/(\varepsilon\kappa)=10$). To obtain
Figure~\ref{fig:charge-Andelman}B from
Figure~\ref{fig:charge-Andelman}A, it is necessary to shift the curves
by an amount indicated in Figure~\ref{fig:charge-Andelman}A by the
horizontal arrows. Since when $x\to-\infty$ the Debye--H\"uckel
electric field increases exponentially
$-\psi'_{\text{DH}}(x)\sim\exp(-x)$ whereas the electric field of
Ref.~\cite{Andelman} increases only linearly $-\psi'(x)\sim -x$, the
curve for that last theory should be shifted to the right much more
that the one for Debye--H\"uckel theory. This makes that the behavior
at $x\to\infty$ is $\psi_{\text{DH}}(x)\sim
4\pi\sigma/(\varepsilon\kappa) \exp(-x)$ and $\psi(x)\sim
4\pi\sigma_{\text{eff}}/(\varepsilon\kappa)\exp(-x)$ with an effective
charge larger than the bare charge $\sigma_{\text{eff}}>\sigma$.  }
\end{figure}

In this case where it is possible to
have finite counterion density when $\psi\to\infty$ and
as discussed in the last section, the function
$F(\psi)$ defined in Eq.~(\ref{eq:function-F}) behaves as $\psi^{1/2}$
as $\psi$ diverges. More precisely, we have here
\begin{equation}
\label{eq:F-BAO-psi-infinity}
F(\psi)
\underset{\psi\to\infty}{=}
\left(\frac{a^3}{2\pi|q|}\right)^{1/2} \psi^{1/2}
+
\mathcal{O}(1)
\end{equation}
so that
\begin{equation}
F^{-1}(-x)\sim\frac{2\pi |q|}{a^3}\, x^2
\end{equation}
when $x\to-\infty$. On the other hand, one recovers Debye--H\"uckel
theory for small potentials. This means that $F^{-1}(-x)\sim A
\exp(-\kappa x)$ for $x\to\infty$ with $A$ an arbitrary constant since
$F$ is determined up to an additive constant.  Suppose that initially
the plane has a small bare charge density $\sigma^{(0)}$. For
Debye--H\"uckel theory the electric potential reads
$\psi_{\text{DH}}(x)=(4\pi\sigma^{(0)}/\varepsilon\kappa)\exp(-\kappa
x)$, so that $\varepsilon \kappa A/(4 \pi)$ may be considered as the
effective charge describing the far field of the plane. As explained
before, $\psi(x)=F^{-1}[-(x+x_0)]$ where the constant of integration
$x_0$ can be determined graphically. Let us choose
$A=4\pi\sigma^{(0)}/(\kappa\varepsilon)$, so that plots of
$F^{-1}(-x)$ and $\psi_{\text{DH}}(x)$ superimpose when $x\to\infty$
(see Figure~\ref{fig:charge-Andelman}).  Now, consider a large bare
charge density $\sigma$ of the plane. To obtain from the curve of
$F^{-1}(-x)$ the correct plot of $\psi(x)$ we should change the origin
of the $x$-axis such that the value of
$dF^{-1}(-x)/dx=-4\pi\sigma/\varepsilon$ at this new origin. Let us
suppose $\sigma$ large enough so that this new origin is in the region
$x\to-\infty$ where $F^{-1}(-x)$ behaves as $x^{2}$. As $\sigma$
increases, we have to shift the origin further to the left or
equivalently to translate the curve of $F^{-1}(-x)$ to the right such
that at the origin, both curves of $F^{-1}(-x)$ and $\psi_{\text{DH}}$
have the same slope. But since $-\psi'_{\text{DH}}(x)$ behaves as
$\exp(-\kappa x)$ which has a faster increase as $x\to -\infty$ than
$-{F^{-1}}'(-x)$ ---which behaves as $-x$--- it is clear that as
$\sigma$ increases we need to shift more to right the curve of
$F^{-1}(-x)$ than the one for $\psi_{\text{DH}}(x)$. On the other hand
for $x\to\infty$, we have $\psi(x)=B\exp(-\kappa x)$ with $B \propto
\exp(-\kappa x_0)$. Since the translation of the curve of $F^{-1}(-x)$
becomes larger and larger than the one for $\psi_{\text{DH}}(x)$, this
means that the factor $B$ increases faster than the bare charge
$\sigma$. The corresponding effective charge therefore increases
faster than $\sigma$ as $\sigma$ is raised. We have also checked this
feature from a direct numerical solution of Eq.~(\ref{eq:andel}). The
above argument may be rationalized, as shown in appendix~\ref{app:c}.

%
%
%
%

In the cases of non-saturation, generalizing the argument presented
before, if the faster divergent counterion density, $n_{\gamma}$,
behaves slower than $\psi$ when $\psi\to\infty$, then the theory will
predict an effective charge that diverges faster than the bare
charge. If, on the other hand, $n_{\gamma}$ diverges faster than
$\psi$ ---but slower than $\psi^{\nu}$ for any $\nu>1$, so there is no
saturation of the effective charge--- then the effective charge will
diverge slower than the bare charge.\footnote{The first example of
footnote~\ref{footnote:cas-limites} is a case where the divergence of
the effective charge will be slower than the bare charge.  }

To sum up, a theory such as that of Ref.~\cite{Andelman} cures a
deficiency of Poisson--Boltzmann theory by setting an upper limit for
the density of micro-ions, at the expense of producing unphysical
effective charges when the non-linear character of the equation
prevails (only for small electrostatic couplings would this theory
give reasonable effective charges, that would then coincide with the
bare charge of the plane since the theory would reduce to
Debye--H\"uckel).  The same would happen for any DFT with the hard
core part included through a local free energy density (one may wish
for instance to use Carnahan--Starling like expressions to account for
excluded volume \cite{Liquids}): such approaches indeed lead to a
divergence of the free energy density in the vicinity of close
packing, so that $F(\psi) \propto \psi^{1/2}$ for large
$\psi$.  Since we have to expect that effective charges are bounded
from above when the bare charge becomes large \cite{Groot}, the above
analysis seems to show that when improving upon Poisson--Boltzmann
theory, incorporation of steric effects through a local functional
and/or neglect of micro-ionic electrostatic correlations is
insufficient.


\section{Summary and discussion}
\label{sec:conclusion}

We have found a necessary and sufficient condition that a generic
local density functional theory should obey to describe the effect of
electrostatic potential saturation for a charged planar
interface. Namely, if in the local relation between the potential and
the densities it is possible to have $\psi\to\infty$ and all densities
finite, there will be no saturation. On the other hand, if as
$\psi\to\infty$ the highest valency counterions density diverges as
$\psi^{\nu}$ for some $\nu>1$ or faster, the theory will predict a
potential saturation. In the other case where the increase of the
density is as $\psi$ or slower, there will not be an electric
potential saturation, hence no charge saturation. For the limiting
cases where the density diverges faster than $\psi$ but slower than
$\psi^{\nu}$ for any $\nu>1$ one should consider the general
condition: $[-\sum_{\alpha} G_{\alpha}(q_{\alpha} \psi)]^{-1/2}$, with
$G_{\alpha}$ defined in Eq.~(\ref{eq:def-G_alpha}), should be
integrable as $\psi\to\infty$ to ensure that the electric potential
saturates.

A natural continuation to this work would be to address the same
question for a colloidal object of any shape. We expect that, for
large colloids, the results found here will hold. However for small
charged objects immersed in an electrolyte, the question remains
open~\cite{Ramanathan}.  It would also be interesting to explore in
more detail the behaviour of effective charges in the framework of
other theories, for instance non-local theories.

At this point, it seems appropriate to come back to the validity of
one of the simplest approaches encompassed by our formalism:
Poisson--Boltzmann (PB) theory. In particular, it seems that the
validity of the saturation regime within PB theory is sometimes
confuse in the literature, which we might trace back to Onsager who
wrote in a seminal paper \cite{Onsager}
\begin{quote}
{\em As soon as the higher order [\/{\rm non-linear\/}] terms in the
Poisson--Boltzmann equation become important, we can no longer expect
the ionic atmosphere to be additive, and then the Poisson--Boltzmann
equation itself becomes unreliable.}
\end{quote} 
This statement (irrelevance of PB as soon as non-linear behaviour sets
in) is justified for simple electrolytes, the system studied by
Onsager in \cite{Onsager}. However, it turns incorrect when there is a
large size and charge asymmetry between the constituents of a charged
mixture, which is the case for colloidal suspensions \cite{Crit}.
This appears in the analytical work of Netz in the no salt limit
\cite{Netz}.  Indeed, PB neglects correlations between microscopic
species, that become prevalent at high electrostatic couplings,
i.e.~when $\Gamma = (q^3 \sigma l_B^2)^{1/2}$ becomes large
\cite{Rouzina,Levin,Netz,Crit}.  On the other hand, PB enters the
saturation regime when the coupling between the charged plane and the
small ions becomes large, which may still be compatible with the
neglect of micro-ionic correlations, provided $\kappa l_B \ll 1$, as
may be seen replacing the saturation value $\sigma_{\text{sat}}
\propto \kappa/l_B$ in the definition of $\Gamma$.  A similar
conclusion is reached enforcing that the correlational contribution to
the pressure in the bulk of the electrolyte (the celebrated term in
$-\kappa^3 k_B T$ appearing within Debye-H\"uckel theory of
electrolytes \cite{McQuarrie,Levin}) should be negligible compared to
the ideal gas contribution $n_b k_BT$: $\kappa^3 \ll n_b$ imposes
$\kappa l_B \ll 1$.


\begin{appendix}
\section{}
\label{sec:appendix}
We prove here that condition~(\ref{eq:hypothese-stabilite}) is a
consequence of the stability of the theory formulated in the framework
of the local DFT~(\ref{eq:DFT-LDA}). First, note that 
\begin{subequations}
\begin{eqnarray}
\label{eq:quantite-derivee}
\frac{d}{dx}\left[
\sum_{\alpha} q_{\alpha} g_{\alpha}(q_{\alpha}\psi(x)) \psi'(x)
\right]=\\
=
\sum_{\alpha} q_{\alpha} n_{\alpha}' \psi' + 
\sum_{\alpha} q_{\alpha} n_{\alpha} \psi'' \\
=-
\sum_{\alpha,\gamma} 
\frac{\partial^2 f}{\partial n_{\alpha}\partial n_{\gamma}} 
n_{\alpha}' n_{\gamma}'  
-\frac{4\pi}{\varepsilon}\rho^2
\label{eq:noyau}
\end{eqnarray}
\end{subequations}
where we have used the minimization equation~(\ref{eq:minimization})
and Poisson equation~(\ref{eq:Poisson}).  If $(\partial^2 f/\partial
n_{\alpha} \partial n_{\gamma})$ is positive definite, this latter
expression~(\ref{eq:quantite-derivee}) is negative and therefore
\begin{equation}
\label{eq:quantite}
\sum_{\alpha} q_{\alpha} g_{\alpha}(q_{\alpha}\psi(x)) \psi'(x)
\end{equation}
is a decreasing function of $x$. Furthermore, the boundary conditions
at $x\to\infty$ impose that~(\ref{eq:quantite}) vanishes at infinity,
so that $\sum_{\alpha} q_{\alpha} g_{\alpha}(q_{\alpha}\psi(x))
\psi'(x)>0$.

The condition that $\partial^2 f/\partial n_{\alpha}\partial n_\gamma$
is positive definite is a sufficient condition to ensure
that~(\ref{eq:hypothese-stabilite}) is satisfied. Let us remark that
this same condition is sufficient for the non-existence of like-charge
attraction between colloids and the absence of
over-charging~\cite{Trizac-charge-like}. It is also a sufficient
condition for the compressibility of the system to be always positive
within the cell model~\cite{Tellez-Trizac-compress-PB}.  Furthermore,
as will be shown in appendix~\ref{sec:appendix-B}, it is a
sufficient and necessary condition to recover Debye--H\"uckel linear
theory where the electric potential is small.

\section{}
\label{sec:appendix-B}

In this appendix we prove that, far from the charged wall, any local
theory reproduces the results of the linear Debye--H\"uckel theory
provided that the Hessian matrix $\partial^2 f/\partial n_{\alpha}
\partial n_{\gamma}$ is positive definite.

Far from the wall, the densities converge to their bulk values
$n_{\alpha}^b$ and the potential vanishes. Then the minimization
equation~(\ref{eq:minimization}) for $\psi=0$ becomes
$\mu_{\alpha}=(\partial f/\partial n_{\alpha})_{b}$, where the
subscript $b$ means that the partial derivative of $f$ is evaluated at
the bulk values of the densities $n_{\alpha}=n_{\alpha}^b$. This provides
the relationship between the chemical potentials and the bulk
densities.

For small values of the potential, the difference
$n_{\alpha}-n_{\alpha}^b$ is small, then we can expand 
\begin{equation}
\frac{\partial f}{\partial n_{\alpha}}
=
\left(\frac{\partial f}{\partial n_{\alpha}}\right)_b
+
\sum_{\gamma}
(n_{\gamma}-n_{\gamma}^{b})
\left(\frac{\partial^2 f}{\partial n_{\alpha}
\partial n_{\gamma}}\right)_b
+
\cdots
\end{equation}
and replacing into the minimization equation~(\ref{eq:minimization})
this gives
\begin{equation}
\sum_{\gamma}
(n_{\gamma}-n_{\gamma}^{b})
\left(\frac{\partial^2 f}{\partial n_{\alpha}
\partial n_{\gamma}}\right)_b
=-q_{\alpha} \psi
\end{equation}
Provided that the Hessian matrix $\left(\frac{\partial^2 f}{\partial n_{\alpha}
\partial n_{\gamma}}\right)_b$ is invertible we have
\begin{equation}
n_{\alpha}=n_{\alpha}^b-
\sum_{\gamma}
\left(\frac{\partial^2 f}{\partial n_{\alpha}
\partial n_{\gamma}}\right)_b^{-1}
q_\gamma \psi
\end{equation}
where $\left(\frac{\partial^2 f}{\partial n_{\alpha} \partial
n_{\gamma}}\right)_b^{-1}$ denotes the $(\alpha,\gamma)$ matrix
element of the inverse of the Hessian matrix $\left(\frac{\partial^2
f}{\partial n_{\alpha} \partial n_{\gamma}}\right)_b$. 

Replacing into the generalized Poisson--Boltzmann
equation~(\ref{eq:Poisson-Boltzmann-general}) we obtain a
Debye--H\"uckel like equation
\begin{equation}
\psi''(x)=\kappa^2 \psi(x)
\end{equation}
with an inverse Debye length $\kappa^{-1}$ defined by
\begin{equation}
\kappa^2=\frac{4\pi}{\varepsilon}
\sum_{\alpha,\gamma} q_{\alpha} q_{\gamma}
\left(\frac{\partial^2 f}{\partial n_{\alpha}
\partial n_{\gamma}}\right)_b^{-1}
\end{equation}
To recover the results from Debye--H\"uckel theory, whatever the
values of the charge $q_{\alpha}$ might be, it is necessary and
sufficient that the Hessian matrix $\left(\frac{\partial^2 f}{\partial
n_{\alpha} \partial n_{\gamma}}\right)_b$ is positive definite, which
ensures the existence of its inverse and that $\kappa^2>0$.

\section{}
\label{app:c}

In this appendix we prove that the theory presented in
Ref.~\cite{Andelman} predicts an effective charge that diverges
exponentially faster than the bare charge when $\sigma\to\infty$.

The function defined by
Eq.~(\ref{eq:function-F}) for the theory of Ref.~\cite{Andelman} reads
\begin{equation}
\label{eq:function-F-BAO}
F_{\text{BAO}}(\psi)=
\frac{\beta q\zeta^{1/2}}{\sqrt{2}\kappa }
\int^{\psi} \frac{d\phi}{\ln\left[1-\zeta+\zeta\cosh(\beta q
\phi)\right]}
\end{equation}
while the one for Debye--H\"uckel theory is
\begin{equation}
\label{eq:function-F-DH}
F_{\text{DH}}(\psi)=\frac{1}{\kappa}
\ln\left(\psi/\psi^{(0)}\right)
\end{equation}
where in the last equation we have made the particular choice
$\psi=\psi^{(0)}$ of the lower bound of integration in the
definition~(\ref{eq:function-F}) of $F_{\text{DH}}$. Suppose that
$\beta q \psi^{(0)}$ is chosen of order one ---say, for example,
$\beta q \psi^{(0)}=1$. Now we choose the lower bound of integration
in Eq.~(\ref{eq:function-F-BAO}) for $F_{\text{BAO}}$ such that
\begin{equation}
F_{\text{BAO}}(\psi)
\underset{\psi\to 0}{=}
\kappa^{-1} \ln\left(\psi/\psi^{(0)}\right) + o(1)
\end{equation}
that is $F_{\text{BAO}}(\psi)$ behaves as $F_{\text{DH}}(\psi)$ when
$\psi\to0$ up to terms that vanish when $\psi\to 0$. On the other hand
we have the behavior~(\ref{eq:F-BAO-psi-infinity}) for
$F_{\text{BAO}}$ when $\psi\to\infty$. Let us define
$\psi_{\text{bare}}=4\pi\sigma/(\kappa \varepsilon)$ and suppose that
the bare charge $\sigma$ of the plane is large enough such that $\beta
q \psi_{\text{bare}} \gg 1$. For this value $\sigma$ of the charge
density of the plane, we have to find the constants of integration
$x_0^{\text{BAO}}$ and $x_0^{\text{DH}}$ of Eq.~(\ref{eq:x-psi}) for
the theory of Ref.~\cite{Andelman} and for Debye--H\"uckel theory
respectively. We can proceed as follows: we first determine the value
of $\psi^{\text{DH,BAO}}$ such that
$F_{\text{DH,BAO}}(\psi^{\text{DH,BAO}})=-x_0^{\text{DH,BAO}}$, then
we deduce $x_0^{\text{DH,BAO}}$. For Debye--H\"uckel theory, applying
the boundary condition~(\ref{eq:BC-psi-x=0}) we have
\begin{equation}
\left.
\frac{d\psi}{dx}
\right|_{x=0}
=
\left.
-\left(\frac{dF_{\text{DH}}}{d\psi}\right)^{-1}
\right|_{\psi=\psi^{\text{DH}}}
=
-\kappa \psi^{\text{DH}} =-\kappa \psi_{\text{bare}}
\end{equation}
Then for $\psi^{\text{DH}}=\psi_{\text{bare}}$ we have
$F_{\text{DH}}(\psi^{\text{DH}})=-x_0^{\text{DH}}$ and therefore $-\kappa
x_0^{\text{DH}}=\ln\left(\psi_{\text{bare}}/\psi^{(0)}\right)$.  For
the theory of Ref.~\cite{Andelman} the boundary
condition~(\ref{eq:BC-psi-x=0}) yields
\begin{eqnarray}
\left.
\frac{d\psi}{dx}
\right|_{x=0}
&=&
\left.
-\left(\frac{dF_{\text{BAO}}}{d\psi}\right)^{-1}
\right|_{\psi=\psi^{\text{BAO}}}
\nonumber\\
&\underset{\beta q\psi^{\text{BAO}}\gg 1}{\sim}&
-\frac{2\kappa}{\sqrt{2\zeta \beta q}}\sqrt{\psi^{\text{BAO}}}
\\
&=&-\kappa \psi_{\text{bare}}
\nonumber
\end{eqnarray}
Therefore we have $\psi^{\text{BAO}}\sim \zeta \beta q
\psi_{\text{bare}}^2/2$ and $-\kappa x_0^{\text{BAO}}=\kappa
F_{\text{BAO}}(\psi^{\text{BAO}})\sim \zeta\beta
q\psi_{\text{bare}}$. Clearly we have $x_0^{\text{BAO}} \gg
x_0^{\text{DH}}$ as $\psi_{\text{bare}}\to\infty$. Furthermore these
constants of integration manifest in the large-$x$ behavior of
$\psi(x)$ as $\psi(x)\sim \psi^{(0)} e^{-\kappa x_0} e^{-\kappa x}$
giving, for $x\to\infty$,
\begin{equation}
\psi_{\text{DH}}(x)= \psi_{\text{bare}} e^{-\kappa x}
\end{equation}
for Debye--H\"uckel theory (as expected), while
\begin{equation}
\psi_{\text{BAO}}(x) 
\underset{x\to\infty}{\sim}
\psi^{(0)} 
e^{\zeta\beta q \psi_{\text{bare}}+o(\psi_{\text{bare}} ) }
e^{-\kappa x}
\end{equation}
for the theory of Ref.~\cite{Andelman}. This theory predicts an
effective charge that diverges exponentially fast when the bare charge
$\sigma\to\infty$.
\end{appendix}

\begin{acknowledgments}
We thank L. Bocquet, Y. Levin and M. Aubouy for useful discussions.
G.~T.~wishes to thank the LPT Orsay, where this work was completed, for
its hospitality. This work was supported by ECOS
Nord/COLCIENCIAS-ICETEX-ICFES action C00P02 of French and Colombian
cooperation.
\end{acknowledgments}

\end{document}